\def\wha{$W_{\rm H\alpha}$}
\def\ha{H$\alpha$}

\documentclass[iop,appendixfloats,numberedappendix]{emulateapj}

\shorttitle{Galaxy Structure vs. Spectral Type}
\shortauthors{Yano et al.}
\usepackage{natbib}
\slugcomment{Accepted for publication in ApJL}

\begin{document}
  
\title{The Relation between Galaxy Structure and Spectral Type: \\ Implications for the Buildup of the Quiescent Galaxy Population at $0.5<\lowercase{z}<2.0$}

\author{Michael Yano\altaffilmark{1}, Mariska Kriek\altaffilmark{1}, Arjen van der Wel\altaffilmark{2}, \& Katherine E. Whitaker\altaffilmark{3,4}}

\altaffiltext{1}{Department of Astronomy, University of California, Berkeley, CA 94720, USA}
\altaffiltext{2}{Max-Planck Institut f\"ur Astronomie, K\"onigstuhl 17, D-69117, Heidelberg, Germany}
\altaffiltext{3}{Department of Astronomy, University of Massachusetts, Amherst, MA 01003, USA}
\altaffiltext{4}{Hubble Fellow}

\begin{abstract}
We present the relation between galaxy structure and spectral type,
using a $K$-selected galaxy sample at $0.5<z<2.0$. Based on
similarities between the UV-to-NIR spectral energy distributions, we
classify galaxies into 32 spectral types. The different types span a
wide range in evolutionary phases, and thus --~in combination with
available CANDELS/F160W imaging~-- are ideal to study the structural
evolution of galaxies. Effective radii ($R_e$) and S\'ersic parameters
($n$) have been measured for 572 individual galaxies, and for each type, we determine
$R_e$ at fixed stellar mass by correcting for the mass-size
relation. We use the rest-frame $U-V$ vs. $V-J$ diagram to investigate
evolutionary trends. When moving into the direction perpendicular to
the star-forming sequence, in which we see the H$\alpha$ equivalent
width and the specific star formation rate (sSFR) decrease, we find a
decrease in $R_e$ and an increase in $n$. On the quiescent sequence we
find an opposite trend, with older redder galaxies being larger. When
splitting the sample into redshift bins, we find that young
post-starburst galaxies are most prevalent at $z>1.5$ and
significantly smaller than all other galaxy types at the same
redshift. This result suggests that the suppression of star formation
may be associated with significant structural evolution at $z>1.5$. At
$z<1$, galaxy types with intermediate sSFRs
($10^{-11.5}-10^{-10.5}~\textrm{yr}^{-1}$) do not have
post-starburst SED shapes. These galaxies have similar sizes as older quiescent
galaxies, implying that they can passively evolve onto the
quiescent sequence, without increasing the average size of the
quiescent galaxy population.
\end{abstract}


\section{INTRODUCTION}\label{sec:int}

One of the most remarkable recent discoveries in extragalactic
astronomy is the finding that galaxies were more compact and denser at
earlier times \citep[e.g.,][]{RWilliams2010,AvanderWel2014}. This
effect is largest for quiescent galaxies, with a factor of 4-5
difference in size between similar-mass galaxies at $z\sim2$ and
$z\sim0$ \citep[e.g.,][]{edaddi2005,pvandokkum2008}. This result poses
a great challenge, as quiescent galaxies are presumably done forming
new stars. Two popular competing theories explaining the size difference
between the distant compact galaxies and the much larger present-day
early-type galaxies are inside-out growth by minor mergers
\citep[e.g.,][]{TNaab2009,RBezanson2009,PHopkins2009}, and quenching
of larger star-forming galaxies at later time \citep[e.g.,][]{MCarollo2013}. 

Yet another puzzling aspect of the population of compact quiescent
galaxies is the nature of their star-forming progenitors. Theoretical
studies predicting the evolutionary tracks of individual galaxies
propose various mechanisms to form compact spheroids, ranging from
gradual shrinking due to violent disk instabilities associated with
intense gas in-streaming and wet minor
mergers \citep[e.g.,][]{ADekel2014,DCeverino2015}, 
centrally-concentrated starbursts triggered by gas-rich major
mergers \citep[e.g.,][]{PHopkins2008,SWellons2015}, or early
assembly in a much denser
universe \citep[e.g.,][]{TNaab2009,SWellons2015}. All of these processes
predict different properties for the direct progenitors of $z\sim2$
compact quiescent galaxies, and thus it is not evident how they can be
identified in observational
studies \citep[e.g.,][]{GBarro2014,ENelson2014,CWilliams2014,PvanDokkum2015}.

To constrain the evolution of compact quiescent galaxies, and assess
the different pathways to quiescence, we need to study how galaxy
structures change with evolutionary phase. In most previous studies,
galaxies were simply divided into broad groups, like star-forming and
quiescent \citep[e.g.,][]{MKriek2009a,AvanderWel2014}. Or, when
following the evolution of galaxies at fixed number density, only the
average properties of large samples were
considered \citep[e.g.,][]{PvanDokkum2010}. However, galaxies are much
more diverse, and by averaging over large samples, or dividing into
crude groups, important evolutionary phases may be missed.

\begin{figure*}  
  \begin{center}   
  \includegraphics[width=1.\textwidth]{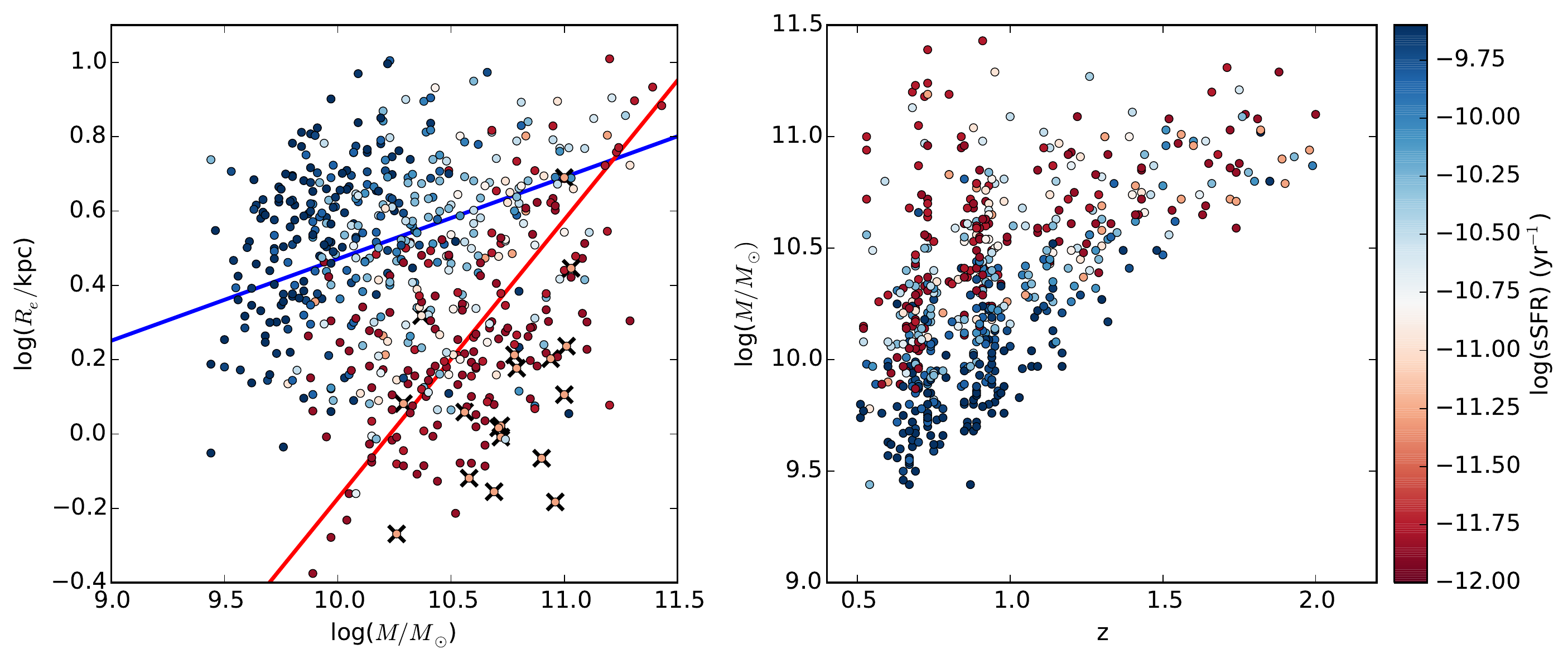} 

  \caption{Effective radius vs. stellar mass (left) and stellar mass
    vs. redshift (right) color coded by sSFR for the individual
    galaxies included in the composite SEDs. In the left panel we show
    the best-fit relations for star-forming (blue line;
    sSFR~$>10^{-11}$yr$^{-1}$) and quiescent (red line;
    sSFR~$<10^{-11}$yr$^{-1}$) galaxies found by adopting the
    power-law index by \citet{AvanderWel2014} of $R_e\propto M^{0.22}$
    and $R_e\propto M^{0.75}$, respectively. In the left panel
    post-starburst galaxies are indicated by black
    crosses.\label{fig:samples}}

  \end{center}
\end{figure*}

In this Letter we use a new approach to study the structural evolution
of galaxies. We have divided a $K$-selected galaxy sample at
$0.5<z<2.0$ into 32 different spectral types \citep{MKriek2011}, using
broad and medium-band photometry from the NEWFIRM Medium-Band
Survey \citep[NMBS;][]{KWhitaker2011}. Part of the NMBS is covered by
 CANDELS \citep{NGrogin2011,AKoekemoer2011}, and thus deep
and high-resolution NIR imaging is available as well. The different
SED types $-$ which span a wide range in evolutionary phases $-$ in
combination with high-resolution rest-frame optical imaging are ideal
for studying the structural evolution of galaxies. 

Throughout this Letter we assume a $\Lambda$CDM cosmology with
$\Omega_{\rm m}=0.3$, $\Omega_{\rm \Lambda}=0.7$, and $H_0=70$ km
s$^{-1}$ Mpc$^{-1}$.

\begin{figure*}
  \begin{center}   
  \includegraphics[width=0.75\textwidth]{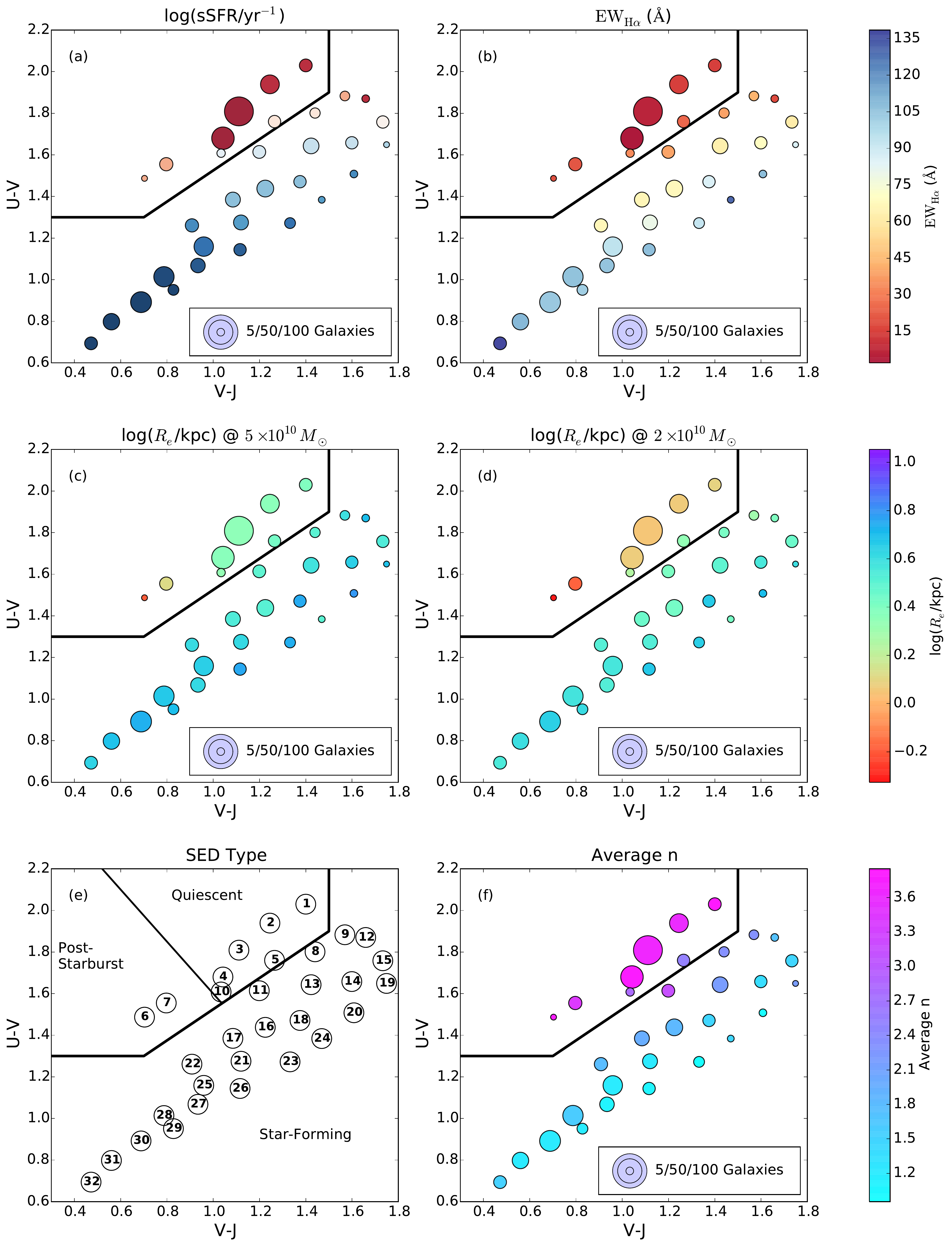} 

  \caption{Rest-frame $U-V$ vs $V-J$ diagrams. The black solid box
      isolates quiescent from star-forming galaxies. Panel (e) further
      shows the distinction between post-starburst and older quiescent
      galaxies in the quiescent box. Each datapoint represents an SED
      type, with the numbers indicated in panel (e). The symbol size
      reflects the number of galaxies per type. Color coding indicates
      sSFR (a, see color bar Figure~\ref{fig:samples}) \ha\
      equivalent width (b) $R_e$ at $5\times10^{10}~M_\odot$ (c) and
      $2\times10^{10}~M_\odot$ (d) and S\'ersic $n$ index (f). This
      figure illustrates the distinct structures of star-forming and
      quiescent galaxies, with $R_e$ decreasing and $n$ increasing
      when moving from the star-forming to the quiescent
      sequence. Furthermore, size increases when moving up the
      quiescent sequence, with post-starburst galaxies being
      significantly smaller than all other galaxy
      types.\label{fig:UVJ_1}}\vspace{-0.1in} 

  \end{center}
\end{figure*}

\section{DATA}\label{sec:data}

We use the composite spectral energy distributions (SEDs)
by \cite{MKriek2011}, which were constructed from multi-wavelength
photometry from the NMBS in the COSMOS field \citep{NScoville2007}. In
summary, $\sim$3500 galaxies with a signal-to-noise ratio of $>$25 in
the K-band were divided into 32 different spectral classes, based on
similarities between their full rest-frame UV-to-NIR SEDs. For each
spectral class we constructed a composite SED by de-redshifting and
scaling the observed photometry of the individual galaxies. The
resulting composite SEDs sample the full $K$-selected galaxy
distribution at $0.5<z<2.0$, and each type presumably represents a
different evolutionary phase.

In previous papers we used this spectral classification method and the
resulting composite SEDs to study star formation and quenching
timescales of galaxies \citep{MKriek2011}, to constrain the shape of
the dust attenuation curve \citep{MKriek2013}, to assess star
formation rate (SFR) indicators \citep{DUtomo2014}, and to study 
X-ray emission as a function of spectral type, stellar mass, and
redshift \citep{TJones2014}. In this Letter we use the different
spectral types to systematically study the structures of galaxies at
$0.5<z<2.0$. 

572 galaxies in our sample are covered by deep {\it Hubble Space
Telescope}/WFC3 imaging as part of CANDELS. For these
galaxies we adopt the effective radii ($R_{\rm e}$;  major axes) and S\'ersic
parameters \citep[$n$;][]{JSersic1968} as measured
by \cite{AvanderWel2012,AvanderWel2014} in the F160W photometric band
using GalFit and Galapagos \citep{CPeng2002,MBarden2012}. The F160W
filter covers rest-frame optical wavelengths for our full redshift
regime. We do not circularize $R_e$ and thus the sizes for
elliptical galaxies may be overestimated. 

Figure~\ref{fig:samples} presents the sizes, stellar masses,
redshifts, and sSFRs of all individual galaxies in our sample. The
sSFRs are derived by fitting the rest-frame UV-to-MIR composite SEDs
with stellar and dust models \citep{DUtomo2014}, and thus are the same
for all galaxies of a given spectral type. Redshifts and stellar
masses are derived using {\tt EAzY} \citep{GBrammer2008} and {\tt
FAST} \citep{MKriek2009b}, respectively, assuming
the \cite{GBruzual2003} stellar population models,
the \cite{GChabrier2003} initial mass function,
the \cite{DCalzetti2000} dust attenuation law, and an exponentially
declining star formation history. Figure~\ref{fig:samples}
illustrates that the targeted mass range changes with
redshift. Furthermore, stellar populations vary with both mass and
redshift. As a consequence, the different sed types will contain
galaxies of different masses and redshifts, and thus caution is
required when interpreting the results.

Figures~\ref{fig:UVJ_1}a and b show the location of the composite SEDs
in the rest-frame $U-V$ vs. $V-J$ (UVJ)
diagram \citep{SWuyts2007,RWilliams2009}, color coded by sSFR and \ha\
equivalent width (\wha), respectively. The rest-frame $U-V$ and $V-J$
colors and \wha\ are directly measured from the composite
SEDs\footnote{\wha\ includes contributions from the blended [N\,{\sc
ii}] and [S\,{\sc ii}] emission lines} \citep{MKriek2011}. Galaxies
show a natural bimodality in the UVJ diagram out to at least
$z\sim2.5$ \citep[e.g.,][]{AMuzzin2013a}, with quiescent and
star-forming galaxies populating two different sequences. The
quiescent sequence is primarily an age
sequence \citep[e.g.,][]{KWhitaker2013}, while the location of
galaxies on the star-forming sequence reflects their dust content and
sSFR. Figures~\ref{fig:UVJ_1}a and b illustrate that the spectral
types comprise quiescent, post-starburst, and star-forming galaxies,
with a range of ages and sSFRs. 

\section{GALAXY STRUCTURE VS. SPECTRAL TYPES}\label{sec:analysis}

\begin{figure*}[t]  
  \begin{center}  
  \includegraphics[width=0.85\textwidth]{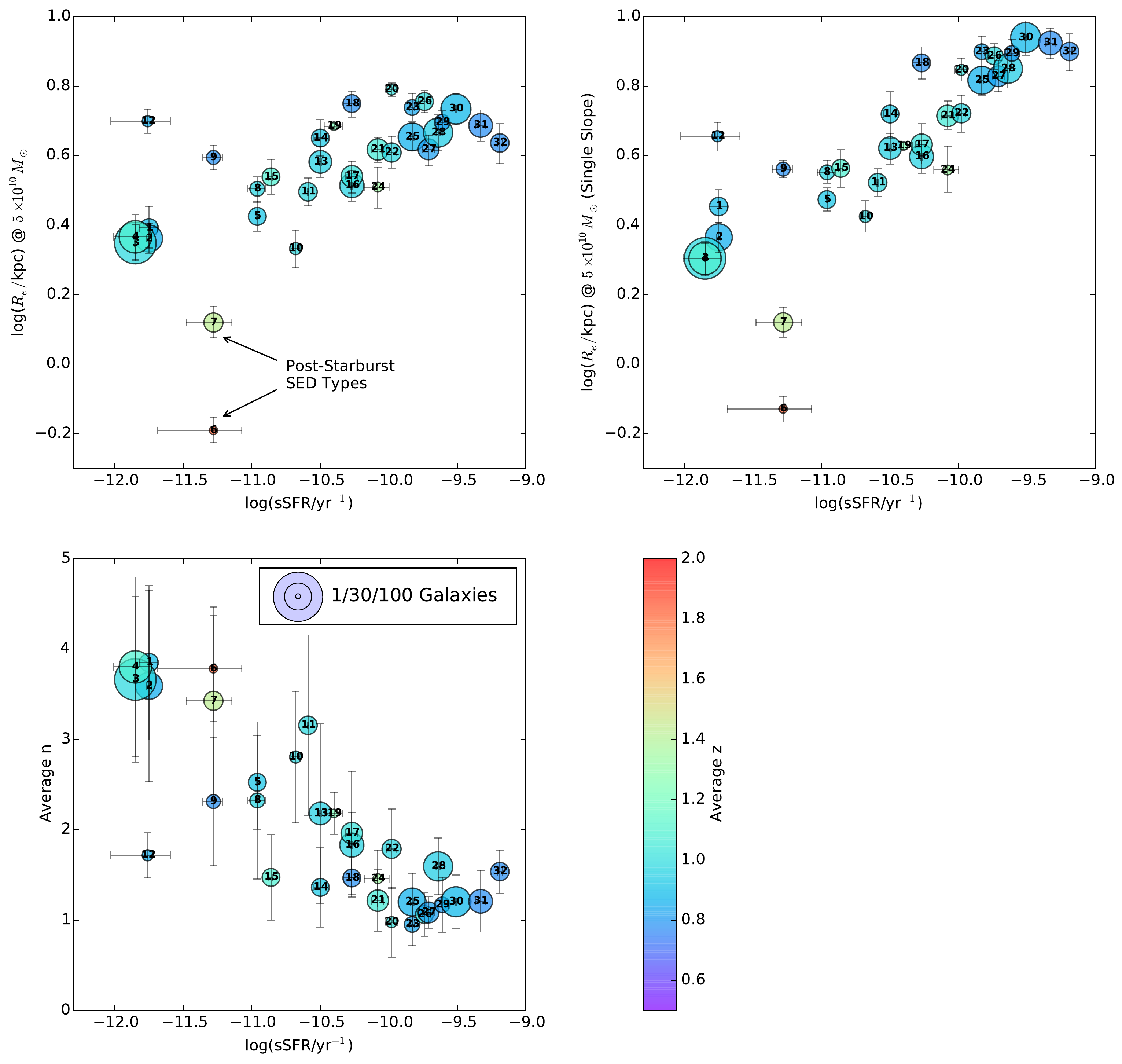}

  \caption{$R_e$ at fixed stellar mass ($5\times10^{10} M_{\odot}$)
    (top panels) and S\'ersic parameter $n$ (bottom) vs. sSFR for
    each SED type.  In the top-left panel $R_e$ is calculated using
    different slopes for the $R_e-M$ relation for quiescent and
    star-forming galaxies, while in the top-right panel we use the same
    slope. The size of each datapoint scales with the number of
    galaxies contained in each type and the color coding reflects the
    mean redshift.  For star-forming galaxies (sSFR~$>10^{-11}$\,yr$^{-1}$)
    $R_e$ gradually decreases with decreasing sSFR, while for
    quiescent galaxies (sSFR~$<10^{-11}$\,yr$^{-1}$) we find an opposite
    trend, with post-starburst SED types 6 and 7 being much smaller
    than older quiescent SED types. The S\'ersic parameter gradually
    increases -- from $n\sim1$ to $n\sim4$ -- with decreasing
    sSFR.\label{fig:SSFR_Re}}

  \end{center}
\end{figure*}

We measure the typical size ($R_e$) for each spectral type at a fixed
stellar mass ($M$), by correcting for the $R_e-M$ relation using a
least-squares fit. Due to incompleteness effects and the small sample
size for some spectral types, we do not constrain the power-law slope,
but fix it to the values found by \cite{AvanderWel2014} for a much
larger and complete sample of galaxies. For quiescent galaxies
(sSFR~$< 10^{-11}\,\rm yr^{-1}$) we assume $R_e\propto M^{0.75}$ and
for star-forming galaxies (sSFR~$> 10^{-11}\,\rm yr^{-1}$) we assume
$R_e\propto M^{0.22}$ (see lines in Fig.~\ref{fig:samples}). To
facilitate comparison with \cite{AvanderWel2014}, we correct all sizes
to a stellar mass of $5\times10^{10}~M_\odot$. To assess the effect of
this mass choice, we also calculate the sizes for a stellar mass of
$2\times10^{10}~M_\odot$. In Figure~\ref{fig:UVJ_1}c and d we color
code the UVJ diagram by $R_e$ at $5\times10^{10}~M_\odot$ and
$2\times10^{10}~M_\odot$, respectively. 

In the top-left panel of Figure~\ref{fig:SSFR_Re} we show $R_e$ at
$5\times10^{10}~M_\odot$ for each SED type as a function of sSFR. The
error bars on $R_e$ indicate the central 68\% of values obtained
through bootstrapping with 2000 iterations. As the slope for
SED types with intermediate sSFRs ($\sim10^{-11}~\textrm{yr}^{-1}$) is
arbitrarily defined, we also show $R_e$ at $5\times10^{10}~M_\odot$
corrected using a single slope of $R_e\propto M^{0.49}$ in the
top-right panel. Finally, we measure the average S\'ersic parameter
for each spectral type, which is presented in UVJ space in
Figure~\ref{fig:UVJ_1}f, and as a function of sSFR in
Figure~\ref{fig:SSFR_Re}. The errors on $n$ present the median
absolute deviation of the values in the sample.

Both figures clearly illustrate that, consistent with previous
studies, star-forming galaxies are larger and have lower S\'ersic
indices than quiescent galaxies of similar mass. The structures of
star-forming galaxies do not change much when we move up the
star-forming sequence in the UVJ diagram. However, when moving into
the direction perpendicular to the star-forming sequence, in which
both \wha\ and sSFR decrease, we see a decrease in $R_e$ and an
increase in $n$ \citep[see also][Whitaker et
al. 2015]{SWuyts2011b}. These trends are also visible in the top
panels in Figure~\ref{fig:SSFR_Re}. The slope of the
$R_e$-sSFR relation becomes steeper if we compare at lower stellar
masses or using a single slope, but the general trends stay the
same. Along the quiescent sequence we observe an increase in $R_e$
when going to redder colors and thus older ages, with post-starburst
galaxies being the smallest. There is no obvious trend between $n$ and
the location on the quiescent sequence.

\begin{figure}  
  \begin{center}  
  \includegraphics[width=0.5\textwidth]{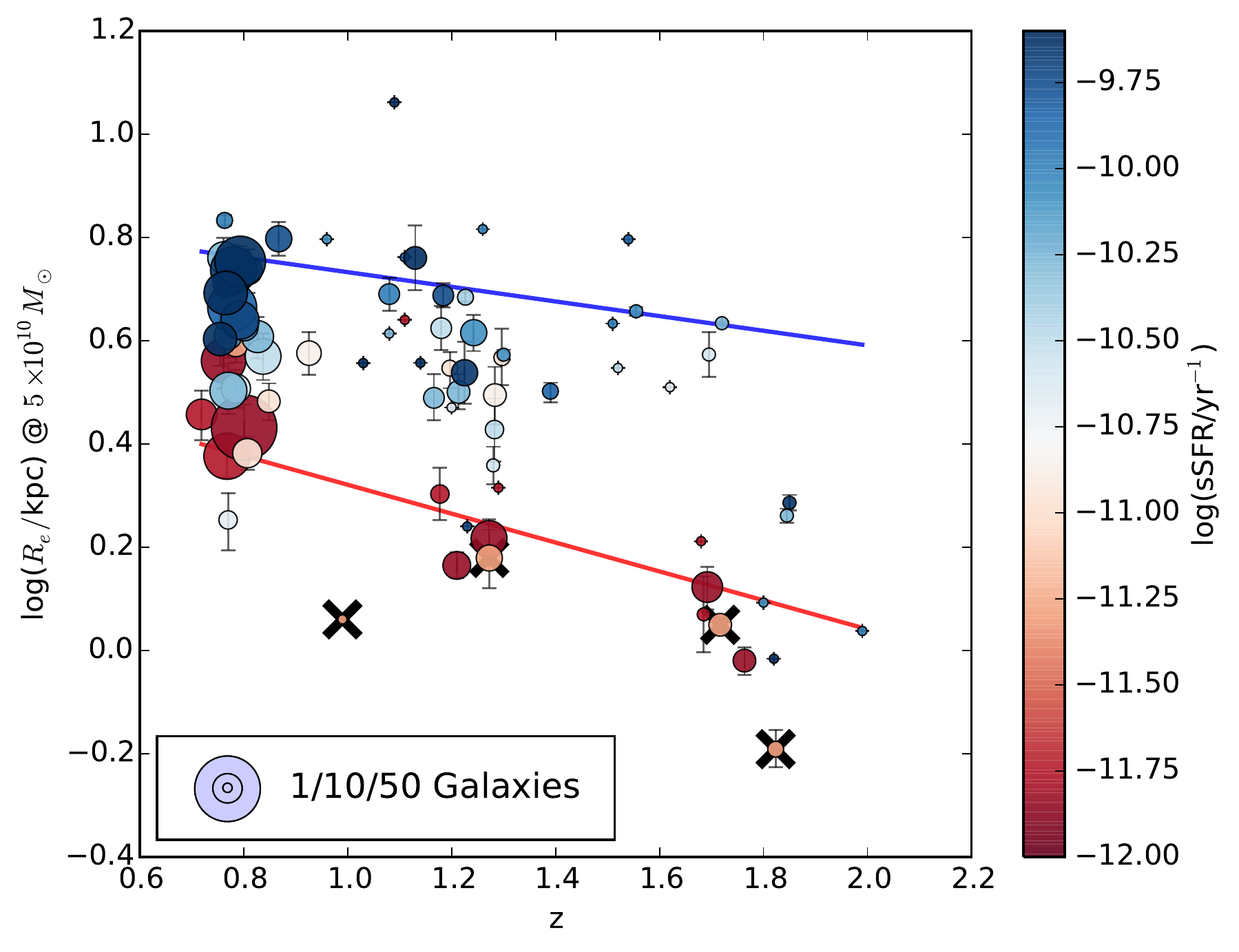}

  \caption{$R_e$ at $5\times10^{10}~M_\odot$ vs. redshift. The samples
	belonging to each SED type are split into three redshift bins,
	the symbol size represents the number of galaxies in each
	datapoint, and the color coding reflects the average
	sSFR. Post-starburst SED types are indicated by the black
	crosses. We do not show vertical error bars for datapoints that
	contain only one galaxy. The lines are the best-fit relations
	for quiescent (sSFR$\,<10^{-11}\rm\,yr^{-1}$) and star-forming
	(sSFR$\,>10^{-11}\rm\,yr^{-1}$) galaxies
	by \cite{AvanderWel2014} for the same stellar mass. At all
	redshifts, star-forming galaxies are larger than quiescent
	galaxies, and both populations increase in size over cosmic
	time.\label{fig:SSFR_Re_dz}}

  \end{center}
\end{figure}

\begin{figure*}  
  \begin{center}  
    \includegraphics[width=1.\textwidth]{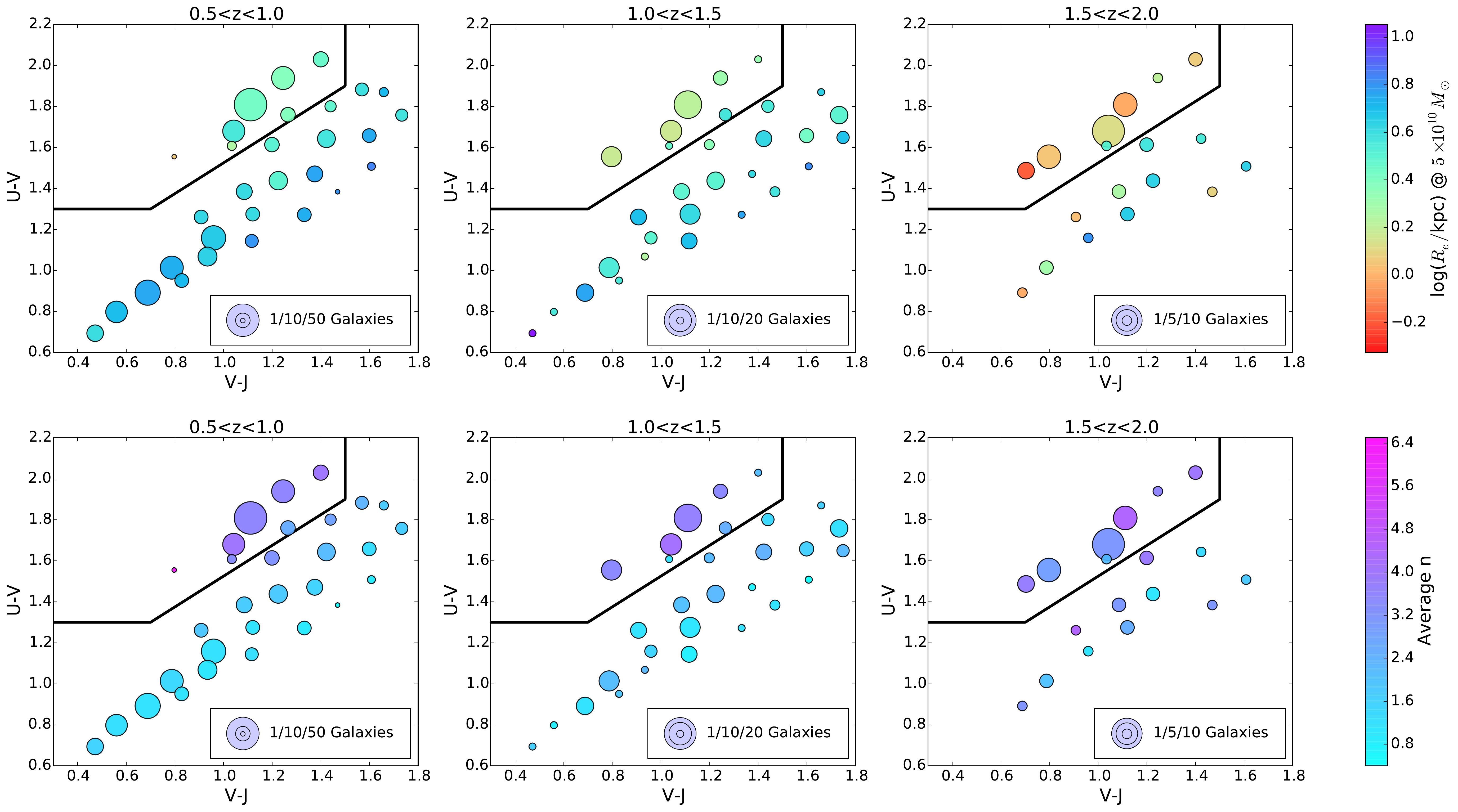} 

    \caption{Rest-frame $U-V$ vs $V-J$ diagrams, color coded by $R_e$
      at $5\times 10^{10}\,M_\odot$ (top panels) and S\'ersic ($n$)
      parameter (bottom panels), with each column representing a
      different redshift bin. Post-starburst galaxies primarily exist
      beyond $z=1$. SED type 6, the youngest post-starburst type, is
      only found at $z>1.5$ in our sample, and significantly more
      compact than all other galaxies at the same
      redshift. Post-starburst SED type 7 is slightly older and more
      similar in size to quiescent galaxies. The galaxies in between
      the star-forming and quiescent sequences have similar ($z<1$) or
      slightly larger sizes ($z>1$) than quiescent
      galaxies.\label{fig:UVJ2}} 

  \end{center}
\end{figure*}

Our composite SEDs include galaxies over a large redshift range, and
differences in the average redshift of the various spectral classes
may contribute to the observed trends. The color coding by redshift in
Figure~\ref{fig:SSFR_Re} indeed indicates that SED type 6 has a higher
average redshift compared to other SED types. To further unravel the
correlation between $R_e$ and sSFR, we split each spectral class into
three different redshift intervals in Figures~\ref{fig:SSFR_Re_dz}
and \ref{fig:UVJ2}.

Consistent with previous studies, Figure~\ref{fig:SSFR_Re_dz}
illustrates that both star-forming and quiescent galaxies were smaller
at earlier times. The size difference between
star-forming and quiescent galaxies of the same mass decreases with
time and is only about $\sim$0.2~dex at $0.5<z<1.0$. This trend
was observed as well by \cite{AvanderWel2014}, as shown by the solid
lines in Figure~\ref{fig:SSFR_Re_dz}. There are small differences
between the two studies, which could be explained by different mass
limits of the galaxy samples, incompleteness effect of our sample (see
Section~\ref{sec:discussion}), and the M/L ratio gradient corrections
applied in \cite{AvanderWel2014}.

Figure~\ref{fig:SSFR_Re_dz} illustrates that the higher average
redshift of post-starburst galaxies indeed contributes to
the smaller size of this type in Figures~\ref{fig:UVJ_1}
and \ref{fig:SSFR_Re}. Nonetheless, Figure
\ref{fig:UVJ2} shows that even in the high-redshift bin
young post-starburst galaxies are significantly smaller than older quiescent
galaxies. Our results support the findings by \cite{KWhitaker2012}
based on ground-based morphological measurements, that post-starburst
galaxies have similar sizes, and perhaps are smaller than older
quiescent galaxies at $z\sim2$.

\section{THE BUILDUP OF THE QUIESCENT SEQUENCE}\label{sec:transitional}

In the previous section we found that young post-starburst galaxies are
smaller than older quiescent types of similar mass at $1.5<z<2.0$.
\cite{KWhitaker2012} used the finding that young quiescent galaxies
are as small as older quiescent galaxies to argue that the addition of
larger recently quenched galaxies cannot explain the size increase of
the quiescent galaxy population. However,
\cite{SBelli2014} argue that there are multiple pathways to quench a
galaxy, and not all quiescent galaxies go through the post-starburst
phase, associated with a short star-formation timescale \citep[see
also][]{GBarro2013,DMarchesini2014,CPapovich2015}. Figure~\ref{fig:UVJ_1} indeed
shows that possible transitional galaxy types with intermediate sSFRs
($10^{-11.5}-10^{-10.5}$~yr$^{-1}$) exist either on the blue end of
the quiescent sequence (types 6 and 7) or in between the star-forming
and quiescent sequences (types 5, 8-12). 

At $1.5<z<2.0$, for which our selection targets galaxies $\gtrsim
10^{10.8}~M_\odot$, the majority of massive galaxies with intermediate
sSFRs are post-starburst galaxies. In fact, all our young and small
post-starburst galaxies (type 6) fall in this redshift range. When
combined with the larger sizes of their older counterparts (type 7)
and other intermediate galaxy types (type 10 and 11), the net size
change due to the addition of new quiescent galaxies will presumably
be small. At $1.0<z<1.5$ the intermediate types have similar or
slightly larger sizes than the quiescent galaxies, possibly leading to
a mild increase of the average size of quiescent galaxies. At
$0.5<z<1.0$ there are no indications that intermediate SED types can
further increase the average size of the quiescent galaxy population,
as they have similar sizes. Post-starburst galaxies are extremely rare
at these low redshifts.

Thus, consistent with the work by \cite{SBelli2014}, our results
suggest that progenitor bias may contribute to the size evolution of
quiescent galaxies at $z>1$, but other mechanisms are needed as
well. Inside-out growth by minor mergers is another popular
explanation for the size growth of quiescent galaxies. We previously
mentioned that there is a gradual size increase of quiescent galaxies
along the quiescent sequence, and thus with age. This trend could be
explained by minor mergers, as older galaxies, which are generally
also the most massive \citep[e.g.,][]{dthomas2005}, may have
experienced more minor mergers. 

In addition to the size growth, our work also gives clues to the
quenching mechanism of galaxies. The similar sizes and S\'ersic
indices of quiescent and intermediate galaxy types at low
redshift suggest that the buildup of the quiescent sequence at
$0.5<z<1.0$ is not associated with much structural change. At
$1.0<z<2.0$ the post-starburst phase becomes more important to the
build up of the quiescent sequence. The small sizes of the young
post-starburst galaxies compared to similar mass galaxies with
slightly higher sSFRs suggest significant structural evolution, which
could either be explained by centrally-concentrated starbursts, possibly
triggered by gas-rich major mergers, or by gradual shrinking due to
violent disk instabilities. Hence, the small sizes of the young
post-starburst galaxies seem inconsistent with the suggested
``passive'' evolutionary tracks by \cite{PvanDokkum2015}. However, we
note that our galaxy sample in the higher redshift bin only consists
of 46 galaxies, and thus larger galaxy samples are needed to confirm
these results.

\section{DISCUSSION}\label{sec:discussion}

In this Letter we study the $HST$/F160W structures of $0.5<z<2.0$
galaxies as a function of SED type. We divided galaxies into different
spectral types, based on their rest-frame UV-to-NIR SEDs. This
approach has several advantages compared to previous studies. First,
we probe a wide range of galaxy evolutionary phases, which allows us
to isolate specific stages. For example, previous work based on
ground-based data showed that young quiescent galaxies are as small as
older quiescent galaxies \citep{KWhitaker2011}. Our more detailed
division shows that the youngest post-starburst galaxies are
significantly smaller than older post-starburst and quiescent galaxies
at $z>1.5$. In addition, we also observe a trend along the quiescent
sequence, with the oldest and reddest SED type being largest. A second
advantage of the composite SEDs is that they are of much higher
quality than individual SEDs, resulting in more accurate fundamental
properties to characterize the evolutionary phase (i.e., \wha\ and
sSFR). Third, as galaxies are matched by their stellar continuum
emission, we automatically exclude galaxies with significant
contributions from active galactic nuclei to their rest-frame optical light,
which could affect the structural measurements. 

However, there are several caveats to our composite SED method as
well. First, the K-band signal-to-noise limit used to select our
sample may introduce a bias toward more compact galaxies. This bias
will primarily affect bins with only few galaxies near the S/N limit
and may explain the small sizes for some of the high-redshift
star-forming bins that consist of only 1 or 2 compact star-forming
galaxies (as the larger galaxies may have been
missed). Second, this study suffers from incompleteness
effects because of an evolving mass limit. We attempt to address the
difference in mass by correcting for the $R_e-M$ relation. However, we
adopt only two slopes, for quiescent and star-forming galaxies. Given
that different types may have different slopes, this correction may
bias our results. We assess this issue by assuming the same slope for
all spectral types and find qualitatively similar
results. Nonetheless, deeper galaxy samples are required to measure
the $R_e-M$ relation for each type to the same mass limit and mitigate
these issues.

Furthermore, we made two major assumptions, which will be addressed in
future work. First, we have only considered F160W sizes, and assumed
that there are no mass-to-light ratio gradients. Nonetheless, we know
that this is incorrect, and that mass sizes are on average 25\%
smaller than rest-frame optical half-light
radii \citep{DSzomoru2013}. As this correction does not correlate with
either stellar mass, sSFR, $R_e$, or $n$, this effect should not
affect the large trends in this work. However, the scatter in the
corrections are large, and systematic trends for specific types may
exist. Second, we compare galaxies at fixed mass. In order to
reconstruct the structural evolution of galaxies, in future studies we
will use the mass profiles for each type, and connect the different
types while taking into account mass growth across redshift.

Finally, as our sample relies on relatively shallow data from the NMBS
in the CANDELS-COSMOS field, we only have few galaxies at higher
redshift. With deeper medium-band photometry from ZFOURGE (I. Labb\'e
et al., in preparation) in 3 CANDELS fields this project can be
extended  using larger samples, to higher redshift, and to lower
masses. 

\acknowledgements  We thank the referee for a constructive report and the NMBS, CANDELS, and 
COSMOS teams for releasing their multi-wavelength datasets to the
community. This work is funded by grant AR-12847, provided by
NASA though a grant from the Space Telescope Science Institute (STScI)
and by NASA-ADAP grant NNX14AR86G. KEW acknowledges support by NASA
through Hubble Fellowship grant \#HF2-51368 awarded by the STScI. The
STScI is operated by the Association of Universities for Research in
Astronomy, Incorporated, under NASA contract NAS5-26555.

%\bibliography{mybib}

\begin{thebibliography}{}
\expandafter\ifx\csname natexlab\endcsname\relax\def\natexlab#1{#1}\fi

\bibitem[{{Barden} {et~al.}(2012){Barden}, {H{\"a}u{\ss}ler}, {Peng},
  {McIntosh}, \& {Guo}}]{MBarden2012}
{Barden}, M., {H{\"a}u{\ss}ler}, B., {Peng}, C.~Y., {McIntosh}, D.~H., \&
  {Guo}, Y. 2012, \mnras, 422, 449

\bibitem[{{Barro} {et~al.}(2013){Barro}, {Faber}, {P{\'e}rez-Gonz{\'a}lez},
  {Koo}, {Williams}, {Kocevski}, {Trump}, {Mozena}, {McGrath}, {van der Wel},
  {Wuyts}, {Bell}, {Croton}, {Ceverino}, {Dekel}, {Ashby}, {Cheung},
  {Ferguson}, {Fontana}, {Fang}, {Giavalisco}, {Grogin}, {Guo}, {Hathi},
  {Hopkins}, {Huang}, {Koekemoer}, {Kartaltepe}, {Lee}, {Newman}, {Porter},
  {Primack}, {Ryan}, {Rosario}, {Somerville}, {Salvato}, \& {Hsu}}]{GBarro2013}
{Barro}, G., {Faber}, S.~M., {P{\'e}rez-Gonz{\'a}lez}, P.~G., {et~al.} 2013,
  \apj, 765, 104

\bibitem[{{Barro} {et~al.}(2014){Barro}, {Trump}, {Koo}, {Dekel}, {Kassin},
  {Kocevski}, {Faber}, {van der Wel}, {Guo}, {P{\'e}rez-Gonz{\'a}lez},
  {Toloba}, {Fang}, {Pacifici}, {Simons}, {Campbell}, {Ceverino},
  {Finkelstein}, {Goodrich}, {Kassis}, {Koekemoer}, {Konidaris}, {Livermore},
  {Lyke}, {Mobasher}, {Nayyeri}, {Peth}, {Primack}, {Rizzi}, {Somerville},
  {Wirth}, \& {Zolotov}}]{GBarro2014}
{Barro}, G., {Trump}, J.~R., {Koo}, D.~C., {et~al.} 2014, \apj, 795, 145

\bibitem[{{Belli} {et~al.}(2014){Belli}, {Newman}, {Ellis}, \&
  {Konidaris}}]{SBelli2014}
{Belli}, S., {Newman}, A.~B., {Ellis}, R.~S., \& {Konidaris}, N.~P. 2014,
  \apjl, 788, L29

\bibitem[{{Bezanson} {et~al.}(2009){Bezanson}, {van Dokkum}, {Tal},
  {Marchesini}, {Kriek}, {Franx}, \& {Coppi}}]{RBezanson2009}
{Bezanson}, R., {van Dokkum}, P.~G., {Tal}, T., {et~al.} 2009, \apj, 697, 1290

\bibitem[{{Brammer} {et~al.}(2008){Brammer}, {van Dokkum}, \&
  {Coppi}}]{GBrammer2008}
{Brammer}, G.~B., {van Dokkum}, P.~G., \& {Coppi}, P. 2008, \apj, 686, 1503

\bibitem[{{Bruzual} \& {Charlot}(2003)}]{GBruzual2003}
{Bruzual}, G., \& {Charlot}, S. 2003, \mnras, 344, 1000

\bibitem[{{Calzetti} {et~al.}(2000){Calzetti}, {Armus}, {Bohlin}, {Kinney},
  {Koornneef}, \& {Storchi-Bergmann}}]{DCalzetti2000}
{Calzetti}, D., {Armus}, L., {Bohlin}, R.~C., {et~al.} 2000, \apj, 533, 682

\bibitem[{{Carollo} {et~al.}(2013){Carollo}, {Bschorr}, {Renzini}, {Lilly},
  {Capak}, {Cibinel}, {Ilbert}, {Onodera}, {Scoville}, {Cameron}, {Mobasher},
  {Sanders}, \& {Taniguchi}}]{MCarollo2013}
{Carollo}, C.~M., {Bschorr}, T.~J., {Renzini}, A., {et~al.} 2013, \apj, 773,
  112

\bibitem[{{Ceverino} {et~al.}(2015){Ceverino}, {Dekel}, {Tweed}, \&
  {Primack}}]{DCeverino2015}
{Ceverino}, D., {Dekel}, A., {Tweed}, D., \& {Primack}, J. 2015, \mnras, 447,
  3291

\bibitem[{{Chabrier}(2003)}]{GChabrier2003}
{Chabrier}, G. 2003, \pasp, 115, 763

\bibitem[{{Daddi} {et~al.}(2005){Daddi}, {Renzini}, {Pirzkal}, {Cimatti},
  {Malhotra}, {Stiavelli}, {Xu}, {Pasquali}, {Rhoads}, {Brusa}, {di Serego
  Alighieri}, {Ferguson}, {Koekemoer}, {Moustakas}, {Panagia}, \&
  {Windhorst}}]{edaddi2005}
{Daddi}, E., {Renzini}, A., {Pirzkal}, N., {et~al.} 2005, \apj, 626, 680

\bibitem[{{Dekel} \& {Burkert}(2014)}]{ADekel2014}
{Dekel}, A., \& {Burkert}, A. 2014, \mnras, 438, 1870

\bibitem[{{Grogin} {et~al.}(2011){Grogin}, {Kocevski}, {Faber}, {Ferguson},
  {Koekemoer}, {Riess}, {Acquaviva}, {Alexander}, {Almaini}, {Ashby}, {Barden},
  {Bell}, {Bournaud}, {Brown}, {Caputi}, {Casertano}, {Cassata}, {Castellano},
  {Challis}, {Chary}, {Cheung}, {Cirasuolo}, {Conselice}, {Roshan Cooray},
  {Croton}, {Daddi}, {Dahlen}, {Dav{\'e}}, {de Mello}, {Dekel}, {Dickinson},
  {Dolch}, {Donley}, {Dunlop}, {Dutton}, {Elbaz}, {Fazio}, {Filippenko},
  {Finkelstein}, {Fontana}, {Gardner}, {Garnavich}, {Gawiser}, {Giavalisco},
  {Grazian}, {Guo}, {Hathi}, {H{\"a}ussler}, {Hopkins}, {Huang}, {Huang},
  {Jha}, {Kartaltepe}, {Kirshner}, {Koo}, {Lai}, {Lee}, {Li}, {Lotz}, {Lucas},
  {Madau}, {McCarthy}, {McGrath}, {McIntosh}, {McLure}, {Mobasher},
  {Moustakas}, {Mozena}, {Nandra}, {Newman}, {Niemi}, {Noeske}, {Papovich},
  {Pentericci}, {Pope}, {Primack}, {Rajan}, {Ravindranath}, {Reddy}, {Renzini},
  {Rix}, {Robaina}, {Rodney}, {Rosario}, {Rosati}, {Salimbeni}, {Scarlata},
  {Siana}, {Simard}, {Smidt}, {Somerville}, {Spinrad}, {Straughn}, {Strolger},
  {Telford}, {Teplitz}, {Trump}, {van der Wel}, {Villforth}, {Wechsler},
  {Weiner}, {Wiklind}, {Wild}, {Wilson}, {Wuyts}, {Yan}, \&
  {Yun}}]{NGrogin2011}
{Grogin}, N.~A., {Kocevski}, D.~D., {Faber}, S.~M., {et~al.} 2011, \apjs, 197,
  35

\bibitem[{{Hopkins} {et~al.}(2009){Hopkins}, {Bundy}, {Murray}, {Quataert},
  {Lauer}, \& {Ma}}]{PHopkins2009}
{Hopkins}, P.~F., {Bundy}, K., {Murray}, N., {et~al.} 2009, \mnras, 398, 898

\bibitem[{{Hopkins} {et~al.}(2008){Hopkins}, {Hernquist}, {Cox}, {Dutta}, \&
  {Rothberg}}]{PHopkins2008}
{Hopkins}, P.~F., {Hernquist}, L., {Cox}, T.~J., {Dutta}, S.~N., \& {Rothberg},
  B. 2008, \apj, 679, 156

\bibitem[{{Jones} {et~al.}(2014){Jones}, {Kriek}, {van Dokkum}, {Brammer},
  {Franx}, {Greene}, {Labb{\'e}}, \& {Whitaker}}]{TJones2014}
{Jones}, T.~M., {Kriek}, M., {van Dokkum}, P.~G., {et~al.} 2014, \apj, 783, 25

\bibitem[{{Koekemoer} {et~al.}(2011){Koekemoer}, {Faber}, {Ferguson}, {Grogin},
  {Kocevski}, {Koo}, {Lai}, {Lotz}, {Lucas}, {McGrath}, {Ogaz}, {Rajan},
  {Riess}, {Rodney}, {Strolger}, {Casertano}, {Castellano}, {Dahlen},
  {Dickinson}, {Dolch}, {Fontana}, {Giavalisco}, {Grazian}, {Guo}, {Hathi},
  {Huang}, {van der Wel}, {Yan}, {Acquaviva}, {Alexander}, {Almaini}, {Ashby},
  {Barden}, {Bell}, {Bournaud}, {Brown}, {Caputi}, {Cassata}, {Challis},
  {Chary}, {Cheung}, {Cirasuolo}, {Conselice}, {Roshan Cooray}, {Croton},
  {Daddi}, {Dav{\'e}}, {de Mello}, {de Ravel}, {Dekel}, {Donley}, {Dunlop},
  {Dutton}, {Elbaz}, {Fazio}, {Filippenko}, {Finkelstein}, {Frazer}, {Gardner},
  {Garnavich}, {Gawiser}, {Gruetzbauch}, {Hartley}, {H{\"a}ussler},
  {Herrington}, {Hopkins}, {Huang}, {Jha}, {Johnson}, {Kartaltepe},
  {Khostovan}, {Kirshner}, {Lani}, {Lee}, {Li}, {Madau}, {McCarthy},
  {McIntosh}, {McLure}, {McPartland}, {Mobasher}, {Moreira}, {Mortlock},
  {Moustakas}, {Mozena}, {Nandra}, {Newman}, {Nielsen}, {Niemi}, {Noeske},
  {Papovich}, {Pentericci}, {Pope}, {Primack}, {Ravindranath}, {Reddy},
  {Renzini}, {Rix}, {Robaina}, {Rosario}, {Rosati}, {Salimbeni}, {Scarlata},
  {Siana}, {Simard}, {Smidt}, {Snyder}, {Somerville}, {Spinrad}, {Straughn},
  {Telford}, {Teplitz}, {Trump}, {Vargas}, {Villforth}, {Wagner}, {Wandro},
  {Wechsler}, {Weiner}, {Wiklind}, {Wild}, {Wilson}, {Wuyts}, \&
  {Yun}}]{AKoekemoer2011}
{Koekemoer}, A.~M., {Faber}, S.~M., {Ferguson}, H.~C., {et~al.} 2011, \apjs,
  197, 36

\bibitem[{{Kriek} \& {Conroy}(2013)}]{MKriek2013}
{Kriek}, M., \& {Conroy}, C. 2013, \apjl, 775, L16

\bibitem[{{Kriek} {et~al.}(2009{\natexlab{a}}){Kriek}, {van Dokkum}, {Franx},
  {Illingworth}, \& {Magee}}]{MKriek2009a}
{Kriek}, M., {van Dokkum}, P.~G., {Franx}, M., {Illingworth}, G.~D., \&
  {Magee}, D.~K. 2009{\natexlab{a}}, \apjl, 705, L71

\bibitem[{{Kriek} {et~al.}(2009{\natexlab{b}}){Kriek}, {van Dokkum},
  {Labb{\'e}}, {Franx}, {Illingworth}, {Marchesini}, \& {Quadri}}]{MKriek2009b}
{Kriek}, M., {van Dokkum}, P.~G., {Labb{\'e}}, I., {et~al.} 2009{\natexlab{b}},
  \apj, 700, 221

\bibitem[{{Kriek} {et~al.}(2011){Kriek}, {van Dokkum}, {Whitaker}, {Labb{\'e}},
  {Franx}, \& {Brammer}}]{MKriek2011}
{Kriek}, M., {van Dokkum}, P.~G., {Whitaker}, K.~E., {et~al.} 2011, \apj, 743,
  168

\bibitem[{{Marchesini} {et~al.}(2014){Marchesini}, {Muzzin}, {Stefanon},
  {Franx}, {Brammer}, {Marsan}, {Vulcani}, {Fynbo}, {Milvang-Jensen}, {Dunlop},
  \& {Buitrago}}]{DMarchesini2014}
{Marchesini}, D., {Muzzin}, A., {Stefanon}, M., {et~al.} 2014, \apj, 794, 65

\bibitem[{{Muzzin} {et~al.}(2013){Muzzin}, {Marchesini}, {Stefanon}, {Franx},
  {Milvang-Jensen}, {Dunlop}, {Fynbo}, {Brammer}, {Labb{\'e}}, \& {van
  Dokkum}}]{AMuzzin2013a}
{Muzzin}, A., {Marchesini}, D., {Stefanon}, M., {et~al.} 2013, \apjs, 206, 8

\bibitem[{{Naab} {et~al.}(2009){Naab}, {Johansson}, \& {Ostriker}}]{TNaab2009}
{Naab}, T., {Johansson}, P.~H., \& {Ostriker}, J.~P. 2009, \apjl, 699, L178

\bibitem[{{Nelson} {et~al.}(2014){Nelson}, {van Dokkum}, {Franx}, {Brammer},
  {Momcheva}, {Schreiber}, {da Cunha}, {Tacconi}, {Bezanson}, {Kirkpatrick},
  {Leja}, {Rix}, {Skelton}, {van der Wel}, {Whitaker}, \&
  {Wuyts}}]{ENelson2014}
{Nelson}, E., {van Dokkum}, P., {Franx}, M., {et~al.} 2014, \nat, 513, 394

\bibitem[{{Papovich} {et~al.}(2015){Papovich}, {Labb{\'e}}, {Quadri}, {Tilvi},
  {Behroozi}, {Bell}, {Glazebrook}, {Spitler}, {Straatman}, {Tran}, {Cowley},
  {Dav{\'e}}, {Dekel}, {Dickinson}, {Ferguson}, {Finkelstein}, {Gawiser},
  {Inami}, {Faber}, {Kacprzak}, {Kawinwanichakij}, {Kocevski}, {Koekemoer},
  {Koo}, {Kurczynski}, {Lotz}, {Lu}, {Lucas}, {McIntosh}, {Mehrtens},
  {Mobasher}, {Monson}, {Morrison}, {Nanayakkara}, {Persson}, {Salmon},
  {Simons}, {Tomczak}, {van Dokkum}, {Weiner}, \& {Willner}}]{CPapovich2015}
{Papovich}, C., {Labb{\'e}}, I., {Quadri}, R., {et~al.} 2015, \apj, 803, 26

\bibitem[{{Peng} {et~al.}(2002){Peng}, {Ho}, {Impey}, \& {Rix}}]{CPeng2002}
{Peng}, C.~Y., {Ho}, L.~C., {Impey}, C.~D., \& {Rix}, H.-W. 2002, \aj, 124, 266

\bibitem[{{Scoville} {et~al.}(2007){Scoville}, {Aussel}, {Brusa}, {Capak},
  {Carollo}, {Elvis}, {Giavalisco}, {Guzzo}, {Hasinger}, {Impey}, {Kneib},
  {LeFevre}, {Lilly}, {Mobasher}, {Renzini}, {Rich}, {Sanders}, {Schinnerer},
  {Schminovich}, {Shopbell}, {Taniguchi}, \& {Tyson}}]{NScoville2007}
{Scoville}, N., {Aussel}, H., {Brusa}, M., {et~al.} 2007, \apjs, 172, 1

\bibitem[{{Sersic}(1968)}]{JSersic1968}
{Sersic}, J.~L. 1968, {Atlas de galaxias australes}

\bibitem[{{Szomoru} {et~al.}(2013){Szomoru}, {Franx}, {van Dokkum}, {Trenti},
  {Illingworth}, {Labb{\'e}}, \& {Oesch}}]{DSzomoru2013}
{Szomoru}, D., {Franx}, M., {van Dokkum}, P.~G., {et~al.} 2013, \apj, 763, 73

\bibitem[{{Thomas} {et~al.}(2005){Thomas}, {Maraston}, {Bender}, \& {Mendes de
  Oliveira}}]{dthomas2005}
{Thomas}, D., {Maraston}, C., {Bender}, R., \& {Mendes de Oliveira}, C. 2005,
  \apj, 621, 673

\bibitem[{{Utomo} {et~al.}(2014){Utomo}, {Kriek}, {Labb{\'e}}, {Conroy}, \&
  {Fumagalli}}]{DUtomo2014}
{Utomo}, D., {Kriek}, M., {Labb{\'e}}, I., {Conroy}, C., \& {Fumagalli}, M.
  2014, \apjl, 783, L30

\bibitem[{{van der Wel} {et~al.}(2012){van der Wel}, {Bell}, {H{\"a}ussler},
  {McGrath}, {Chang}, {Guo}, {McIntosh}, {Rix}, {Barden}, {Cheung}, {Faber},
  {Ferguson}, {Galametz}, {Grogin}, {Hartley}, {Kartaltepe}, {Kocevski},
  {Koekemoer}, {Lotz}, {Mozena}, {Peth}, \& {Peng}}]{AvanderWel2012}
{van der Wel}, A., {Bell}, E.~F., {H{\"a}ussler}, B., {et~al.} 2012, \apjs,
  203, 24

\bibitem[{{van der Wel} {et~al.}(2014){van der Wel}, {Franx}, {van Dokkum},
  {Skelton}, {Momcheva}, {Whitaker}, {Brammer}, {Bell}, {Rix}, {Wuyts},
  {Ferguson}, {Holden}, {Barro}, {Koekemoer}, {Chang}, {McGrath},
  {H{\"a}ussler}, {Dekel}, {Behroozi}, {Fumagalli}, {Leja}, {Lundgren},
  {Maseda}, {Nelson}, {Wake}, {Patel}, {Labb{\'e}}, {Faber}, {Grogin}, \&
  {Kocevski}}]{AvanderWel2014}
{van der Wel}, A., {Franx}, M., {van Dokkum}, P.~G., {et~al.} 2014, \apj, 788,
  28

\bibitem[{{van Dokkum} {et~al.}(2008){van Dokkum}, {Franx}, {Kriek}, {Holden},
  {Illingworth}, {Magee}, {Bouwens}, {Marchesini}, {Quadri}, {Rudnick},
  {Taylor}, \& {Toft}}]{pvandokkum2008}
{van Dokkum}, P.~G., {Franx}, M., {Kriek}, M., {et~al.} 2008, \apjl, 677, L5

\bibitem[{{van Dokkum} {et~al.}(2010){van Dokkum}, {Whitaker}, {Brammer},
  {Franx}, {Kriek}, {Labb{\'e}}, {Marchesini}, {Quadri}, {Bezanson},
  {Illingworth}, {Muzzin}, {Rudnick}, {Tal}, \& {Wake}}]{PvanDokkum2010}
{van Dokkum}, P.~G., {Whitaker}, K.~E., {Brammer}, G., {et~al.} 2010, \apj,
  709, 1018

\bibitem[{{van Dokkum} {et~al.}(2015){van Dokkum}, {Nelson}, {Franx}, {Oesch},
  {Momcheva}, {Brammer}, {F{\"o}rster Schreiber}, {Skelton}, {Whitaker}, {van
  der Wel}, {Bezanson}, {Fumagalli}, {Illingworth}, {Kriek}, {Leja}, \&
  {Wuyts}}]{PvanDokkum2015}
{van Dokkum}, P.~G., {Nelson}, E.~J., {Franx}, M., {et~al.} 2015, \apj, 813, 23

\bibitem[{{Wellons} {et~al.}(2015){Wellons}, {Torrey}, {Ma}, {Rodriguez-Gomez},
  {Vogelsberger}, {Kriek}, {van Dokkum}, {Nelson}, {Genel}, {Pillepich},
  {Springel}, {Sijacki}, {Snyder}, {Nelson}, {Sales}, \&
  {Hernquist}}]{SWellons2015}
{Wellons}, S., {Torrey}, P., {Ma}, C.-P., {et~al.} 2015, \mnras, 449, 361

\bibitem[{{Whitaker} {et~al.}(2012){Whitaker}, {Kriek}, {van Dokkum},
  {Bezanson}, {Brammer}, {Franx}, \& {Labb{\'e}}}]{KWhitaker2012}
{Whitaker}, K.~E., {Kriek}, M., {van Dokkum}, P.~G., {et~al.} 2012, \apj, 745,
  179

\bibitem[{{Whitaker} {et~al.}(2011){Whitaker}, {Labb{\'e}}, {van Dokkum},
  {Brammer}, {Kriek}, {Marchesini}, {Quadri}, {Franx}, {Muzzin}, {Williams},
  {Bezanson}, {Illingworth}, {Lee}, {Lundgren}, {Nelson}, {Rudnick}, {Tal}, \&
  {Wake}}]{KWhitaker2011}
{Whitaker}, K.~E., {Labb{\'e}}, I., {van Dokkum}, P.~G., {et~al.} 2011, \apj,
  735, 86

\bibitem[{{Whitaker} {et~al.}(2013){Whitaker}, {van Dokkum}, {Brammer},
  {Momcheva}, {Skelton}, {Franx}, {Kriek}, {Labb{\'e}}, {Fumagalli},
  {Lundgren}, {Nelson}, {Patel}, \& {Rix}}]{KWhitaker2013}
{Whitaker}, K.~E., {van Dokkum}, P.~G., {Brammer}, G., {et~al.} 2013, \apjl,
  770, L39

\bibitem[{{Williams} {et~al.}(2014){Williams}, {Giavalisco}, {Cassata},
  {Tundo}, {Wiklind}, {Guo}, {Lee}, {Barro}, {Wuyts}, {Bell}, {Conselice},
  {Dekel}, {Faber}, {Ferguson}, {Grogin}, {Hathi}, {Huang}, {Kocevski},
  {Koekemoer}, {Koo}, {Ravindranath}, \& {Salimbeni}}]{CWilliams2014}
{Williams}, C.~C., {Giavalisco}, M., {Cassata}, P., {et~al.} 2014, \apj, 780, 1

\bibitem[{{Williams} {et~al.}(2009){Williams}, {Quadri}, {Franx}, {van Dokkum},
  \& {Labb{\'e}}}]{RWilliams2009}
{Williams}, R.~J., {Quadri}, R.~F., {Franx}, M., {van Dokkum}, P., \&
  {Labb{\'e}}, I. 2009, \apj, 691, 1879

\bibitem[{{Williams} {et~al.}(2010){Williams}, {Quadri}, {Franx}, {van Dokkum},
  {Toft}, {Kriek}, \& {Labb{\'e}}}]{RWilliams2010}
{Williams}, R.~J., {Quadri}, R.~F., {Franx}, M., {et~al.} 2010, \apj, 713, 738

\bibitem[{{Wuyts} {et~al.}(2007){Wuyts}, {Labb{\'e}}, {Franx}, {Rudnick}, {van
  Dokkum}, {Fazio}, {F{\"o}rster Schreiber}, {Huang}, {Moorwood}, {Rix},
  {R{\"o}ttgering}, \& {van der Werf}}]{SWuyts2007}
{Wuyts}, S., {Labb{\'e}}, I., {Franx}, M., {et~al.} 2007, \apj, 655, 51

\bibitem[{{Wuyts} {et~al.}(2011){Wuyts}, {F{\"o}rster Schreiber}, {van der
  Wel}, {Magnelli}, {Guo}, {Genzel}, {Lutz}, {Aussel}, {Barro}, {Berta},
  {Cava}, {Graci{\'a}-Carpio}, {Hathi}, {Huang}, {Kocevski}, {Koekemoer},
  {Lee}, {Le Floc'h}, {McGrath}, {Nordon}, {Popesso}, {Pozzi}, {Riguccini},
  {Rodighiero}, {Saintonge}, \& {Tacconi}}]{SWuyts2011b}
{Wuyts}, S., {F{\"o}rster Schreiber}, N.~M., {van der Wel}, A., {et~al.} 2011,
  \apj, 742, 96

\end{thebibliography}

\end{document}